\documentclass[12pt]{article}

\usepackage{amssymb}
\usepackage{amsmath}

\makeatletter
\renewcommand{\@oddhead}{\hfil \thepage}
\renewcommand{\@oddfoot}{}
\makeatother

\begin{document}

\thispagestyle{empty}
\renewcommand{\refname}{\normalsize References}

\begin{center}
\textbf{\large On the nature of mass-energy constituents of the universe}\\[0.5cm]
\textsf{V.E. Kuzmichev, V.V. Kuzmichev}\\[0.5cm]
\textit{Bogolyubov Institute for Theoretical Physics, National
Academy of Sciences of Ukraine, Kiev, 03143 Ukraine}
\end{center}

\begin{abstract}
On the basis of our quantum cosmological approach we show that
there can be two previously unknown types of collective states in
the universe. One of them relates to a gravitational field,
another is connected with a matter (scalar) field which fills the
universe on all stages of its evolution. The increase in number of
the quanta of the collective excitations of the gravitational
field manifests itself as an expansion of the universe. The
collective excitations of the scalar field above its true vacuum
reveal themselves mainly in the form of dark matter and energy.
Under the action of the gravitational forces they decay and
produce the non-baryonic dark matter, optically bright and dark
baryons. We have calculated the corresponding energy densities
which prove to be in good agreement with the data from the recent
observations.
\end{abstract}
\ \\

\section*{\normalsize 1. Introduction}
The recent astrophysical observations provide the strong empirical
evidence that the total energy density of the universe equals to
$\Omega_{0} = 1$ to within 10\% (in units of critical density)
\cite{1,2}. At the same time the mean matter density consistent
with big-bang nucleosynthesis is estimated by the value
$\Omega_{M} \approx 0.3$ \cite{2,3,4}. It is assumed that the
density $\Omega_{M}$ is contributed by the optically bright
baryons ($\Omega_{*} \approx 0.003$ \cite{3,Sal1,Sal2}) and dark
matter which consists of the dark baryons ($\Omega_{DB} \approx
0.04$) and matter of uncertain origin and composition (known as
cold dark matter, $\Omega_{CDM} \approx 0.3$). The contribution
from neutrinos $\Omega_{\nu}$ depends on the neutrino mass and its
lower limit is comparable with $\Omega_{*}$ \cite{3,4}. The lack
of density $\Omega_{X} \sim 0.7$, where $\Omega_{X} = \Omega_{0} -
\Omega_{M}$, is ascribed to so-called dark energy \cite{3,4,5,6}.
Its nature is unknown and expected properties are unusual. This
dark energy is unobservable (in no way could it be detected in
galaxies), spatially homogeneous and, as it is expected, it has
large, negative pressure. The last property should guarantee an
agreement with the present-day accelerated expansion of the
universe observed in the type Ia supernova Hubble diagram
\cite{7,8}.

Thus modern cosmology poses the principle question about the
nature of the components of our universe and their percentage in
the total energy density. In this paper we make an attempt to
explain the matter content of our universe on the basis of the
conjuncture that the collective excitations of previously unknown
type exist in it. At the heart of this approach we put the quantum
cosmology formulated in \cite{9,10,11,12,13}. It is shown that
there are two types of collective states in the quantum universe.
One of them relates to a gravitational field, another is connected
with a matter (scalar) field which fills the universe on all
stages of its evolution. The quantization of the fields which
describe the geometrical and matter properties of the universe is
made. We demonstrate that in the early epoch the non-zero energy
of the vacuum in the form of the primordial scalar field is a
source of the transitions of the quantum universe from one state
to another. During these transitions the number of the quanta of
the collective excitations of the gravitational field increases
and this growth manifests itself as an expansion of the universe.
The collective excitations of the scalar field above its true
vacuum reveal themselves mainly in the form of dark (nonluminous)
matter and energy. Under the action of the gravitational forces
they decay and produce the non-baryonic dark matter, optically
bright and dark baryons and leptons. For the state of the universe
with large number of the matter field quanta the total energy
density, density of (both optically bright and dark) baryons, and
density of non-baryonic dark matter are calculated. The
theoretical values prove to be in good agreement with the data
from the recent observations in our universe.

Below we use dimensionless variables. The modified Planck values
are chosen as units of length $l = \sqrt{2 G /(3 \pi)} = 0.74
\times 10^{-33}$ cm, mass/energy $m_{Pl} = \sqrt{3 \pi /(2 G)} =
2.65 \times 10^{19}$ GeV, and time $t_{Pl} = \sqrt{2 G /(3 \pi)} =
2.48 \times 10^{-44}$ s, where $\hbar = c = 1$. The unit of
density is $\rho_{Pl} = 3 /(8 \pi G l^{2}) =
\frac{9}{16}\,\widetilde{\rho}_{Pl}$, where
$\widetilde{\rho}_{Pl}$ is the standard Planck density. The scalar
field $\phi$ is measured in units $\phi_{Pl} = \sqrt{3 /(8 \pi G)}
= l \sqrt{\rho_{Pl}}$, while the potential $V(\phi)$ in
$\rho_{Pl}$.

\section*{\normalsize 2. Quantum cosmology}
\subsection*{\normalsize 2.1 Equation of motion}
Let us consider a homogeneous, isotropic and closed universe
filled with primordial matter in the form of the uniform scalar
field $\phi$. The potential $V(\phi)$ of this field will be
regarded as a positive definite function of $\phi$. The quantum
analog of such universe is described by the equation
\cite{9,10,11,12,13}
\begin{equation}\label{1}
   2\,i\, \partial _{T} \Psi = \left[ \partial _{a}^{2} -
  \frac{2}{a^{2}}\,\partial _{\phi }^{2} - U \right] \Psi \,,
\end{equation}
where its wavefunction $\Psi (a, \phi, T)$ depends on the scale
factor $a$ ($0 \leq a < \infty$), the scalar field $\phi$ ($-
\infty < \phi < \infty$), and the time variable $T$ ($- \infty < T
< \infty$). The last is related to the synchronous proper time $t$
by the differential equation $dt = a\,d T$ \cite{9,10}. The
functional
\begin{eqnarray}\label{2}
    U = a^{2} - a^{4}\,V(\phi )
\end{eqnarray}
plays the role of an effective potential of the interaction
between the gravitational and matter fields. It should be
emphasized that (\ref{1}) is not an ordinary Schr\"{o}dinger
equation with a potential (\ref{2}). It is a constraint equation
for the wavefunction (see e.g. \cite{Dirac,16}). The famous
Wheeler-DeWitt equation for the minisuperspace model (the
wavefunction depends on the finite number of variables) is the
special case ($\partial_{T} \Psi = 0$) of the more general
equation (\ref{1}). The discussion of the possible solutions of
the Wheeler-DeWitt equation, its interpretation, and the
corresponding references can be found in \cite{17,18}. In the
region $a^{2} V > 1$ the solutions of the Wheeler-DeWitt equation
and the solutions of (\ref{1}) as well can be associated with the
exponential expansion of the universe \cite{9,13,22}.

A general solution of (\ref{1}) can be represented by a
superposition
\begin{equation}\label{3}
  \Psi (a, \phi , T) = \int_{- \infty}^{\infty }\!dE\,\mbox{e}^{\frac{i}{2} E T}\,
               C(E)\, \psi _{E}(a, \phi )
\end{equation}
of the states $\psi _{E}$ which satisfy a stationary equation
\begin{eqnarray}\label{4}
 \left( -\,\partial _{a}^{2} + \frac{2}{a^{2}}\,\partial _{\phi }^{2} +
             U - E  \right)  \psi _{E} = 0.
\end{eqnarray}
Here $E$ is the eigenvalue. It has a physical dimension of action,
while in (\ref{4}) it is written in units $\hbar/2$. The function
$C(E)$ characterizes the distribution in $E$ of the states of the
universe at the instant $T = 0$.

Let us compare (\ref{1}) and (\ref{4}) with the corresponding
Schr\"{o}dinger equations. It is well known that when a
Hamiltonian does not depend explicitly on the time variable the
quantum system is described by the states with the time-dependence
in the form $\exp (-i \mathcal{E} t)$. The time variable $t$ and
the classical energy $\mathcal{E}$ can be related with each other
through the formal transformation $\mathcal{E} \rightarrow i\,
\partial _{t}$ \cite{14,15}. In the case of quantum cosmology
there is an analogous formal transformation between the classical
parameter $E$, which enters into the energy-momentum tensor of
radiation
\begin{eqnarray}
  \tilde T^{0}_{0} & = & \frac{E}{a^{4}},\quad
 \tilde T^{1}_{1} = \tilde T^{2}_{2} = \tilde T^{3}_{3} =
 -\,\frac{E}{3\, a^{4}},\nonumber \\
  \tilde T^{\mu }_{\nu } & = & 0
  \ \ \mbox{for} \ \  \mu \neq \nu, \label{5}
\end{eqnarray}
and the time variable $T$, $E \rightarrow -\, 2\, i\, \partial
_{T}$ (minus originates from the specific character of the
gravitational problem, while the additional factor $2$ results
from the choice of units). This analogy demonstrates that the
role, played in ordinary quantum mechanics by the time variable
$t$, is assumed by the variable $T$. Thus in the framework of the
minisuperspace model the equation (\ref{1}) solves the problem of
correct definition of the time coordinate in quantum cosmology.
The role of time in quantum gravity and the difficulties with the
introduction of time variable suitable for the description of the
dynamics in quantum cosmology are elucidated in \cite{16,17}.

The effective potential $U$ has the form of barrier of the finite
width and height in the variable $a$. It is convenient to
represent the solution of (\ref{4}) in the form
\begin{equation}\label{6}
  \psi _{E} (a, \phi) = \int_{- \infty}^{\infty }\!d\epsilon\,
  \varphi _{\epsilon}(a,  \phi )\, f_{\epsilon} (\phi ; E),
\end{equation}
where the function $\varphi_{\epsilon}$ satisfies the equation
\begin{eqnarray}\label{7}
 \left( -\,\partial _{a}^{2} + U - \epsilon \right)  \varphi _{\epsilon} =
 0.
\end{eqnarray}
The eigenvalue $\epsilon$ and eigenfunction $\varphi_{\epsilon}$
depend parametrically on $\phi$. The functions
$\varphi_{\epsilon}$ describe the states of the continuous
spectrum. Using the explicit form of the solution of (\ref{7}) in
the region $a^{2} V \gg 1$ \cite{9,13}, one can obtain the
normalization condition for the functions $\varphi_{\epsilon}$
\begin{eqnarray}\label{8}
    \int_{0}^{\infty}\!da\ \varphi _{\epsilon}^{*} (a, \phi)\,
    \varphi _{\epsilon '} (a, \phi) = \delta (\epsilon - \epsilon
    ').
\end{eqnarray}
Then the function $f_{\epsilon}$ will satisfy the equation
\begin{eqnarray}\label{9}
    \partial _{\phi}^{2} f_{\epsilon} + \int_{-\infty}^{\infty}\!d\epsilon '
    K_{\epsilon \epsilon '}\, f_{\epsilon '} = 0,
\end{eqnarray}
where the kernel is
\begin{eqnarray}
    K_{\epsilon \epsilon '}(\phi; E) & = & \int_{0}^{\infty}\!da\ \varphi _{\epsilon}^{*}\,
    \partial_{\phi}^{2} \varphi_{\epsilon '} + 2\,\int_{0}^{\infty}\!da\
    \varphi _{\epsilon}^{*}\, \partial_{\phi} \varphi_{\epsilon '} \partial_{\phi} \nonumber \\
    & + & \frac{1}{2}\,(\epsilon ' - E)\, \int_{0}^{\infty}\!da\ \varphi _{\epsilon}^{*}\,
     a^{2} \varphi_{\epsilon '}. \label{10}
\end{eqnarray}

This equation can be reduced to a solvable form if one defines
concretely the problem. In order to find the function $\psi_{E}$,
the equation (\ref{4}) must be supplemented with the boundary
condition.

\subsection*{\normalsize 2.2 Scenario of the origin of quantum universe}
From (\ref{4}) and (\ref{7}) it follows that these equations
coincide at $a^{2} \rightarrow \infty$. It means that in the
asymptotic region $a^{2} \gg 1$ the wavefunction $\psi_{E}$ can be
taken in the form of the function $\varphi_{\epsilon}$ (or
superposition of such functions) which in turn can be chosen as
the sum of the wave $\varphi^{(-)}_{\epsilon}$ incident upon the
barrier $U$ and the outgoing wave $\varphi^{(+)}_{\epsilon}$
\cite{13}. The amplitude of the scattered wave has the poles in
the complex plane of $\epsilon$ at $\epsilon = \epsilon_{n} +i
\,\Gamma_{n}$, where $\epsilon_{n}$ and $\Gamma_{n}$ are the
positions and widths of the quasistationary levels with the
numbers $n = 0, 1, 2 \ldots$)

Just as in classical cosmology which uses a model of the slow-roll
scalar field \cite{18,19}, in quantum cosmology based on (\ref{1})
it makes sense to consider a scalar field $\phi$ which slowly
evolves into its true vacuum state, $V(\phi_{vac}) = 0$, from some
initial state $\phi = \phi_{start}$, where $V(\phi_{start}) \sim
\widetilde{\rho}_{Pl}$. The latter condition makes it possible to
consider the evolution of the universe in time in the classical
sense. This allows to describe the process of formation of quantum
universe in time (see below).

For the slow-roll potential $V$, when $\left|d \ln V / d \phi
\right|^{2} \ll 1$, the function $\varphi _{\epsilon }$ describes
the universe in the adiabatic approximation. In this case one can
separate the slow motion (with respect to the variable $\phi$)
from the rapid one (with respect to the variable $a$) in the
universe. This means that the universe expands (or contracts) more
rapidly than the state of matter has time to change.

The calculations demonstrate \cite{9,10,13} that the first ($n =
0$) quasistationary state appears when $V$ decreases to $V_{in} =
0.08 = 0.045 \widetilde{\rho}_{Pl}$. It has the parameters:
$\epsilon_{0} = 2.62 = 1.31 \hbar, \ \Gamma_{0} = 0.31 = 1.25
\times 10^{43}\, \mbox{s}^{-1}$. The lifetime of the universe in
this state, $\tau = 0.8 \times 10^{-43}$ s, is three times greater
than the Planck time $t_{Pl}$. The instant of the origination of
the first quasistationary state can be taken as a reference point
of time against the scale of $T$.

A further decrease of $V$ leads to an increase of the number of
quasistationary states of the universe. The levels which have
emerged earlier shift towards the oscillator values
$\epsilon_{n}^{0} = 4 n + 3$ which they would have in the limit $V
\rightarrow 0$, while their widths
\begin{equation}\label{11}
    \Gamma_{n} \approx \exp \left\{- 2 \!\int_{a_{1}}^{a_{2}} da \sqrt{U -
    \epsilon_{n}}\right\} \quad \mbox{at} \quad \Gamma_{n}\ll \epsilon_{n}
\end{equation}
decrease exponentially. Here $a_{1}$ and $a_{2}$ ($a_{1} < a_{2}$)
are the turning points specified by the condition $U =
\epsilon_{n}$. As a result the lifetime of the universe in such
states is many orders greater than the Planck time and at $V \sim
10^{-122}$ it reaches the values $\tau \sim 10^{61} \sim 10^{17}$
s comparable with the age of our universe.

Thus the whole scenario can be represented in the following way.
In the quantum cosmological system of the most general type which
is described by a superposition of the waves
$\varphi^{(\pm)}_{\epsilon}$ incident upon the barrier and
scattered by the barrier \cite{9,10,13}, it occurs the formation
of the fundamentally new state as a result of a slow decreasing of
the potential $V$ (vacuum energy density \cite{17,18}). When the
potential $V$ reaches the value $V_{in}$ the wave
$\varphi_{\epsilon}$ with $\epsilon \approx \epsilon_{0}$
penetrates into the prebarrier region \cite{13} and it results in
the transition of the cosmological system to a quasistationary
state. In this state the system is characterized by the
expectation values of the scale factor $\langle a \rangle _{n=0}=
1.25 = 0.93 \times 10^{-33} \, \mbox{cm}$ and total energy density
$\langle \rho \rangle_{n=0} = 1.16 = 0.65 \widetilde{\rho}_{Pl} =
3.35 \times 10^{93} \, \mbox{g}/\mbox{cm}^{3}$ \cite{10,13}. These
parameters are of Planck scale. Such a new formation has a
lifetime which exceeds the Planck time and one can consider it as
a quantum cosmological system with well-defined physical
properties. We call it the quantum universe. Such universe can
evolve by means of change of its quantum state. In every quantum
state it can be characterized by the energy density, observed
dimensions, lifetime (age), proper dimensions of non-homogeneities
of the matter density, amplitude of fluctuation of radiation
temperature, power spectrum of density perturbations, angular
structure of the radiation anisotropies, deceleration parameter,
total entropy, and others \cite{13}.

\section*{\normalsize 3. Collective states}
\subsection*{\normalsize 3.1 Collective excitations of the gravitational field}
Within the lifetime the state of the universe can be considered
with a high accuracy as a stationary state which takes the place
of a quasistationary one when its width becomes zero \cite{20}.
Taking into account that such states emerge at $V \ll 1$, while in
the prebarrier region ($a < a_{1}$) $a^{2} V < 1$ always, the
solutions of (\ref{7}) can be written in the form of expansion in
powers of $V$ on the interval $\Delta \phi = |\phi_{vac} -
\phi_{start}|$. We have
\begin{eqnarray}
\varphi_{n} & = & \left.|n\right\rangle
  - \frac{V}{4}\left[\frac{1}{8}\sqrt{N(N-1)(N-2)(N-3)}
\left.|n-2\right\rangle \right.\nonumber \\
  & + & \sqrt{N(N-1)}\left(N- \frac{1}{2}\right)\left.|n-1\right\rangle \nonumber \\
  & - & \sqrt{(N+1)(N+2)}\left(N+\frac{3}{2}\right)\left.|n+1\right\rangle \nonumber \\
  & - & \left. \frac{1}{8}\sqrt{(N+1)(N+2)(N+3)(N+4)}\left.|n+2\right\rangle\right]\nonumber \\
  & - & O(V^{2})
\label{12}
\end{eqnarray}
for the wavefunction and
\begin{eqnarray}\label{13}
\epsilon_{n}  = \epsilon_{n}^{0} - \frac{3}{4}\,V\left[2N(N+1) +
1\right] - O(V^{2})
\end{eqnarray}
for the position of the level, where $N = 2n + 1$. Here $|n
\rangle$ is the eigenfunction, $\epsilon_{n}^{0} = 2 N + 1$ is the
eigenvalue of the equation for an isotropic oscillator with zero
orbital angular momentum
\begin{equation}\label{14}
    \left( - \partial_{a}^{2} + a^{2} - \epsilon_{n}^{0}\right)\left.|n\right\rangle =
    0.
\end{equation}

The wavefunction $\langle a|n \rangle$ describes the geometrical
properties of the universe as a whole. Since the gravitational
field is considered as a variation of space-time metric \cite{21},
then this wavefunction also characterizes the quantum properties
of the gravitational field which is considered as an aggregate of
the elementary excitations of quantum oscillator (\ref{14}). It
should be emphasized that in this approach the role of the dynamic
variables is taken by the variables $(a, \phi)$ of the
minisuperspace and after the quantization such elementary
excitations cannot be identified with gravitons. Introducing the
operators
\begin{equation}\label{15}
A^{\dag}= \frac{1}{\sqrt{2}}\,(a - \partial_{a}), \quad
 A = \frac{1}{\sqrt{2}}\,(a + \partial_{a}),
\end{equation}
the state $|n \rangle$ can be represented in the form
\begin{eqnarray}
 |n \rangle  & = &
 \frac{1}{\sqrt{N!}}\,(A^{\dag})^{N}\,|vac \rangle, \quad
  A\,|vac \rangle  = 0, \nonumber \\
  |vac \rangle & = &
 \left(\frac{4}{\pi}\right)^{1/4}\exp \left\{-
 \frac{a^{2}}{2}\right\}.\label{16}
\end{eqnarray}
Since $A^{\dag}$ and $A$ satisfy the ordinary canonical
commutation relations, $[A, A^{\dag}] = 1,\ [A,A] = [A^{\dag},
A^{\dag}] = 0$, then one can interpret them as the operators for
the creation and annihilation of quanta of the collective
excitations of the gravitational field (in the sense indicated
above). We shall call it g-quantum. The integer $N$ gives the
number of g-quanta in the n-th state of the gravitational field.

Passing in (\ref{14}) to the ordinary physical units it is easy to
see that the states (\ref{16}) can be interpreted as those which
emerge as a result of motion of some imaginary particle with the
mass $m_{Pl}$ and zero orbital angular momentum in the field with
the potential energy $U(R) = k_{Pl} R^{2} /2$, where $R = l a$ is
a ``radius'' of the curved universe, while $k_{Pl}=
(m_{Pl}\,c^{2})^{3}\, (\hbar\,c)^{-2}$ can be called a ``stiffness
coefficient of space (or gravitational field)'', $k_{Pl} = 4.79
\times 10^{85} \, \mbox{GeV}/\mbox{cm}^{2} = 0.76 \times 10^{83}
\, \mbox{g}/\mbox{s}^{2}$ . This motion causes the equidistant
spectrum of energy $E_{n}= \hbar\,\omega_{Pl}$ $\left(N +
\frac{1}{2}\right) = 2\,\hbar\,\omega_{Pl}\left(n +
\frac{3}{4}\right)$, where $\hbar\,\omega_{Pl} = m_{Pl}\,c^{2}$ is
the energy of g-quantum, $\omega_{Pl} = t_{Pl}^{-1}$ is the
oscillation frequency.

The interaction of the gravitational field with a vacuum which has
non-zero energy density ($V(\phi) \neq 0$) in the range $\Delta
\phi$ results in the fact that the wavefunction (\ref{12}) is a
superposition of the states $|n \rangle, \ |n \pm 1\rangle$ and
$|n \pm 2 \rangle$. From (\ref{12}) and (\ref{13}) it follows that
the quantum universe can be characterized by a quantum number $n$.

The physical state $|n \rangle$ (\ref{16}) is chosen so that $|0
\rangle$ describes an initial state of the gravitational field.
From the point of view of the occupation number representation
there is only one g-quantum in such state. When the gravitational
field transits from the state $|n \rangle$ into the neighbouring
one $|n + 1 \rangle$ two g-quanta are created, while the
corresponding energy increases by $2\, \hbar\, \omega_{Pl}$. In
the inverse transition two g-quanta are absorbed and the energy
decreases. The transitions with the creation/absorption of one
g-quantum are forbidden. The vacuum state $|vac \rangle$ in
(\ref{16}) describes the universe without any g-quantum.

The g-quanta are bosons. Since the probability of the creation of
a boson per unit time grows with the increase in number of bosons
in a given state (see e.g. \cite{14}), then an analogous effect
must reveal itself in the quantum universe during the creation of
g-quanta. As the average value is $\langle a \rangle \sim \sqrt{(N
+ 1)/2}$, then the growth in number $N$ of g-quanta (or number of
levels $n$) means the increase in expectation value of the scale
factor. In other words the expansion of the universe reflects the
fact of the creation of g-quanta (under the transition to a higher
level). The increase in probability of the creation of these
quanta per unit time leads to the accelerated expansion of the
universe. This phenomenon becomes appreciable only when $n$
reaches the large values. The observations of type Ia supernovae
provide the evidence that today our universe is expanding with
acceleration \cite{4,7,8}. The above-mentioned qualitative
explanation of this phenomenon is confirmed by the concrete
quantitative calculations of the deceleration parameter \cite{13}.
On this point the theory is in good agreement with the
measurements of type Ia supernovae.

\subsection*{\normalsize 3.2 Collective excitations of the matter field}
When the potential $V$ decreases to the value $V \ll 0.1$ the
number of available states of the universe increases up to $n \gg
1$. By the moment when the scalar field will roll in the location
where $V(\phi_{vac}) = 0$ the universe can be found in the state
with $n \gg 1$. This can occur because the emergence of new
quantum levels and the (exponential) decrease in width of old ones
result in the appearance of competition between the tunneling
through the barrier $U$ and allowed transitions between the
states, $n \rightarrow n, \ n \pm 1, \ n \pm 2$. A comparison
between these processes demonstrates \cite{9,10,13} that the
transition $n \rightarrow n + 1$ with creation of g-quanta is more
probable than any other allowed transitions or decay. The vacuum
energy  of the early universe originally stored by the field
$\phi$ with a potential $V(\phi_{start})$ is a source of
transitions.

According to accepted model the scalar field $\phi$ descends to
the state with zero energy density, $V(\phi_{vac}) = 0$. Then due
to the quantum fluctuations the field $\phi$ begins to oscillate
with a small amplitude about equilibrium vacuum value
$\phi_{vac}$. We shall describe these new states of the quantum
universe assuming that by the moment when the field $\phi$ reaches
the value $\phi_{vac}$ the universe transits to the state with $n
\gg 1$. This means that the ``radius'' of the universe has become
$\langle a \rangle > 10\, l$. The wavefunction of the universe is
$\psi_{E} = \sum _{n} \psi_{n}$, where the function $\psi_{n}$
describes the universe in the state with a given $n$ and takes the
following form up to the terms $\sim O(V^{2})$
\begin{equation}\label{18}
    \psi_{n} (a, \phi) = \langle a | n \rangle \,f_{n}(\phi; E)
    \quad \mbox{at} \quad n \gg 1,
\end{equation}
where the function $f_{n}$ satisfies the equation
\begin{equation}\label{19}
    \left[\partial_{x}^{2} + z - V(x) \right] f_{n} = 0.
\end{equation}
Here $x = \sqrt{m/2}\,(2N)^{3/4}\,(\phi - \phi_{vac})$
characterizes the deviation of the field $\phi$, $z =
(\sqrt{2N}/m)\,\left(1 - E/(2N)\right),$ $V(x) =
(2N)^{3/2}\,V(\phi)/m$, $m$ is a dimensionless parameter which it
is convenient to choose as $m^{2} = \left[\partial_{\phi}^{2}
V(\phi)\right]_{\phi_{vac}}$. The formulae (\ref{18}) and
(\ref{19}) follow from (\ref{6}), (\ref{9}), and (\ref{12}) if one
takes into account resonance behaviour of the wave
$\varphi_{\epsilon}$ at $\epsilon = \epsilon_{n}$ in the
prebarrier region \cite{11,13}.

Since $\langle a \rangle = \sqrt{N/2}$, where averaging was
performed over the states (\ref{18}), then $V(x)$ is a potential
energy of the scalar field contained in the universe with the
volume $\sim \langle a \rangle^{3}$. Expanding $V(x)$ in powers of
$x$ we obtain
\begin{equation}\label{20}
    V(x) = x^{2} + \alpha\,x^{3} + \beta\,x^{4} + \ldots,
\end{equation}
where the parameters are
\begin{equation}\label{21}
    \alpha =
    \frac{\sqrt{2}}{3}\,\frac{\lambda}{m^{5/2}}\,\frac{1}{(2N)^{3/4}}\,,
    \quad
    \beta = \frac{1}{6}\,\frac{\nu}{m^{3}}\,\frac{1}{(2N)^{3/2}}\,,
\end{equation}
and $\lambda = \left[\partial_{\phi}^{3} V(\phi)
\right]_{\phi_{vac}}, \ \nu = \left[\partial_{\phi}^{4} V(\phi)
\right]_{\phi_{vac}}$. Since $N \gg 1$, then the coefficients are
$|\alpha| \ll 1$ and $|\beta| \ll 1$ even at $m^{2} \sim \lambda
\sim \nu$. Therefore (\ref{19}) can be solved using the
perturbation theory for stationary problems with a discrete
spectrum. We take for the state of the unperturbed problem the
state of the harmonic oscillator with the equation of motion
\begin{equation}\label{22}
    \left[\partial_{x}^{2} + z^{0} - x^{2} \right] f_{n}^{0} = 0.
\end{equation}
In the occupation number representation one can write
\begin{eqnarray}
 f_{ns}^{0} & = & \frac{1}{\sqrt{s!}}\,(B_{n}^{\dag})^{s} f_{n0}^{0},
 \quad B_{n}\,f_{n0}^{0} = 0, \nonumber \\
  f_{n0}^{0} (x) & = &
 \left(\frac{1}{\pi}\right)^{1/4}\exp \left\{-
 \frac{x^{2}}{2}\right\}\label{23}
\end{eqnarray}
with $z^{0} = 2s + 1$, where $s = 0, 1, 2 \ldots$, and
\begin{equation}\label{24}
B_{n}^{\dag}= \frac{1}{\sqrt{2}}\,(x - \partial_{x}), \qquad
 B_{n} = \frac{1}{\sqrt{2}}\,(x + \partial_{x}).
\end{equation}
Here $B_{n}^{\dag}$ ($B_{n}$) can be interpreted as the creation
(annihilation) operator which increases (decreases) the number of
quanta of the collective excitations of the scalar field in the
universe in the n-th state. We shall call them the matter quanta.
The variable $s$ is the number of matter quanta in the state $n
\gg 1$. It can be considered as an additional quantum number.

Using (\ref{19}), (\ref{20}) and (\ref{23}) we obtain
\begin{equation}\label{25}
    z = 2 s + 1 + \Delta z,
\end{equation}
where
\begin{eqnarray}
    \Delta z & = & \frac{3}{2}\,\beta \left(s^{2} + s +
    \frac{1}{2}\right) - \frac{15}{8}\,\alpha^{2} \left(s^{2} + s +
    \frac{11}{30}\right) \nonumber \\
    & - & \frac{\beta^{2}}{16} \left(34 s^{3} + 51 s^{2} + 59 s +
    21\right)\label{26}
\end{eqnarray}
takes into account a self-action of matter quanta. In order to
determine the physical meaning of the quantities which enter
(\ref{19}) with the potential (\ref{20}) we rewrite it in the
ordinary physical units,
\begin{eqnarray}
\left( - \frac{\hbar^{2}}{2 \mu}\,\partial_{r}^{2} +
\frac{1}{2}\,\mu\, \omega^{2} r^{2} + {\cal E}_{1}\,\left(
\frac{r}{l}\right)^{3} + {\cal E}_{2}\,\left(
\frac{r}{l}\right)^{4} \right.\nonumber \\
 \left. + \ldots - {\cal E} \right) f_{n}(r) =
0,\label{27}
\end{eqnarray}
where we denote $\mu = m_{Pl}\,m^{-1},\ r = l\,x,\ \omega =
m\,t_{Pl}^{-1},\ {\cal E}_{1} = m \, m_{Pl}\, c^{2}\, \alpha / 2,
\ {\cal E}_{2} = m \, m_{Pl}\, c^{2}\, \beta / 2,$ ${\cal E} = m
\, m_{Pl}\, c^{2}\, z / 2$. Here $l = \sqrt{\hbar/(\mu \omega)}$
is the Planck length. According to (\ref{27}) an imaginary
particle with a mass $\mu$ performs the anharmonic oscillations
and generates the energy spectrum
\begin{equation}\label{28}
    {\cal E} = \hbar\,\omega \left(s + \frac{1}{2} + \frac{\Delta
    z}{2}\right),
\end{equation}
where $\hbar\, \omega = m\,m_{Pl}\,c^{2}$ is the energy of matter
quantum, and $m$ can be interpreted as its mass (in units
$m_{Pl}$). The quantity
\begin{equation}\label{29}
    M = m \,\left(s + \frac{1}{2}\right) + \Delta M,
\end{equation}
where $\Delta M = m\,\Delta z/2$, is a mass of the universe with
$s$ matter quanta. Since we are interested in large values of $s$,
then for the estimation of $\Delta M$ let us assign the values $s
\sim 10^{80}$ and $n \sim 10^{122}$. These parameters describe our
universe, where $s$ is equal to the equivalent number of baryons
in it and $\langle a \rangle \sim 10^{28}$ cm is a size of its
observed part \cite{11,12}. In this case $\Delta M \sim
O\left((\nu/m^{2}) 10^{-24},\ (\lambda/m^{2})^{2}
10^{-24}\right)$. Hence when the number of the matter quanta
becomes very large their self-action can be a fortiori neglected.

\section*{\normalsize 4. The matter content of the quantum universe}
\subsection*{\normalsize 4.1 Einstein-Friedmann equation in terms of the average values}
In the early epoch (on the interval $\Delta \phi$) there is no
matter in the ordinary sense in the universe, since the field
$\phi$ is only a form of existence of the vacuum, while the vacuum
energy density is decreasing with time. In spite of the fact that
$V(\phi_{vac}) = 0$, the average value $\langle V \rangle \neq 0$.
It contributes to the vacuum energy density in the epoch when the
matter quanta are created and determines the cosmological constant
$\Lambda = 3 \, \langle V \rangle$. The universe is filling with
matter in the form of aggregate of quanta of the collective
excitations of the primordial scalar field.

Let us consider the possible matter content of the universe from
the point of view of quantum cosmology. For this purpose we change
from the quantum equation (\ref{4}) to the corresponding Einstein
equation for homogeneous, isotropic, and closed universe filled
with the uniform matter and radiation. Averaging over the states
$\psi_{E}$ we obtain the Einstein-Friedmann equation in terms of
the average values
\begin{equation}\label{30}
   \left(\frac{\partial_{t} \langle a \rangle}{\langle a \rangle}
    \right)^{2} = \langle \rho \rangle - \frac{1}{\langle a
    \rangle^{2}}\,,
\end{equation}
where we have neglected the dispersion, $\langle a^{2} \rangle
\sim \langle a \rangle ^{2}$, and $\langle a^{6} \rangle \sim
\langle a \rangle ^{6}$. The average value
\begin{equation}\label{31}
    \langle \rho \rangle = \langle V \rangle + \frac{2}{\langle a
    \rangle^{6}}\,\langle - \partial_{\phi}^{2} \rangle + \frac{E}{\langle a \rangle^{4}}
\end{equation}
is a total energy density in some fixed instant of time with the
Hubble constant $H_{0} = \partial_{t} \langle a \rangle / \langle
a \rangle$. The first term is the energy density of the vacuum
with the equation of state $p_{v} = - \rho_{v} = - \langle V
\rangle$, the second term is the matter energy density, while the
last describes the contribution of the radiation.

Using the wavefunction (\ref{18}) and passing in (\ref{16}) and
(\ref{23}) to the limit of large quantum numbers, for the energy
densities of the vacuum $\Omega_{v}$ and matter $\Omega_{qm}$ we
obtain
\begin{eqnarray}\label{32}
 \Omega_{v} =  \frac{M}{12\langle a
 \rangle^{3}\,H_{0}^{2}}\,,\quad \Omega_{qm}= \frac{16\, M}{\langle a
 \rangle^{3}\,H_{0}^{2}}\,,
\end{eqnarray}
where $M = m \left(s + \frac{1}{2}\right)$ is the mass of the
universe with ``radius'' $\langle a \rangle = \sqrt{N/2}$. From
the definition of $z$ in (\ref{19}) it follows that $\langle a
\rangle = M + E/(4 \langle a \rangle)$. For the matter-dominant
era, $M \gg E/(4 \langle a \rangle)$, from (\ref{30}) and
(\ref{32}) we find
\begin{equation}\label{33}
    \Omega = 1.066, \quad \Omega_{v} = 0.006, \quad \Omega_{qm} =
    1.060,
\end{equation}
where $\Omega = \langle \rho \rangle H_{0}^{-2}$. From (\ref{33})
it follows that the quantum universe in all states with $n \gg 1$
and $s \gg 1$ looks like spatially flat. A main contribution to
the energy density of the universe is made by the collective
excitations of the scalar field above its true vacuum. The total
density $\Omega$ can be compared with the present-day density of
our universe, $\Omega_{0} = 1 \pm 0.12$ \cite{1} or $\Omega_{0} =
1.025 \pm 0.075$ \cite{2}.

The matter quanta are bosons. They have non-zero mass/energy,
while their state according to (\ref{23}) depends on the state of
the gravitational field. This means that the matter quanta are
subject to the action of gravity. Due to this fact they can decay
into the real particles (e.g. baryons and leptons) that have to be
present in the universe in small amounts (because of the weakness
of the gravitational interaction). The main contribution to the
energy density will still be made by the matter quanta.

\subsection*{\normalsize 4.2 The matter quantum decay}
In order to make a numerical estimate we consider the matter
quantum ($\phi$) decay scheme
\begin{equation}\label{34}
    \phi \rightarrow \phi'\,\nu \,n \rightarrow \phi'\, \nu \, p \, e^{-}\,
    \bar{\nu},
\end{equation}
where $\phi'$ is the quantum  of the residual excitation, which
reveals itself in the universe in the form of the non-baryonic
dark matter. Neutrino $\nu$ takes away the spin. Assuming that the
quanta $\phi$, just as the neutrons $n$, decay independently, we
obtain the law of production of protons in the form
\begin{equation}\label{35}
    s_{p} = s \left[1 + \frac{1}{\Gamma_{n} -
    \Gamma_{\phi}}\left(\Gamma_{\phi}\,\mbox{e}^{-\Gamma_{n}\,\Delta
    t} - \Gamma_{n}\,\mbox{e}^{-\Gamma_{\phi}\,\Delta
    t}\right)\right],
\end{equation}
where $s$ is the number of quanta $\phi$ at some arbitrarily
chosen initial instant of time $t'$, $\Gamma_{n}$ and
$\Gamma_{\phi}$ are the decay rates of neutron and quantum $\phi$
respectively, $\Delta t = t - t'$ is time interval during which
$s_{p}$ protons were produced. Since we are interested in matter
density in the universe today, then for numerical estimations we
choose $\Delta t$ equal to the age of the universe, $\Delta t =
14$ Gyr \cite{3}. For experimentally measured decay rate of
neutron, $\Gamma_{n} = 1.13 \times 10^{-3}\ s^{-1}$ \cite{3} we
have: $\Gamma_{n}\,\Delta t = 5 \times 10^{14} \gg 1$. We shall
suppose that the decay of quantum $\phi$ is caused by the action
of the gravitational forces. This means the following condition
must be fulfilled
\begin{equation}\label{36}
    \Gamma_{\phi} \ll \Gamma_{n}.
\end{equation}
Taking into account these two remarks one can simplify (\ref{35})
\begin{equation}\label{37}
    s_{p} = \bar{s} = s \left[1 - \mbox{e}^{-
    \Gamma_{\phi}\,\Delta t}\right],
\end{equation}
where $\bar{s}$ is an average number of the quanta $\phi$ which
decay during the time interval $\Delta t$. The law of proton
production (\ref{37}) implies that all primordial neutrons have to
decay up to now. The quanta $\phi'$ in (\ref{34}) are assumed to
be the stable particles (with lifetime greater than $\Delta t$),
and their number is equal to $\bar{s}$. The decay rate
$\Gamma_{\phi}$ is unknown and it must be determined theoretically
on the basis of vertex modelling of complex decay (\ref{34}) or
extracted from the astrophysical data. Since, generally speaking,
the decay rate of the quanta $\phi$ may change with time, then by
$\Gamma_{\phi}$ it should be implied the mean probability of decay
on the time interval $\Delta t$,
\begin{equation}\label{38}
    \Gamma_{\phi} = \frac{1}{\Delta t}\int_{t'}^{t}\!
    dt\,\zeta (t)\,\gamma (t),
\end{equation}
where $\gamma (t)$ is the decay rate at the instant of time $t$
when the energy density in the universe equals to $\rho (t)$, and
weight function $\zeta (t)$ takes into account influence of other
accompanying processes on the decay rate (e.g. space-time
curvature).

The density of (optically bright and dark) baryons is
\begin{eqnarray}\label{39}
 \Omega_{B}= \frac{16\, \bar{s}\, m_{p}}{\langle a
 \rangle^{3}\,H_{0}^{2}}\,,
\end{eqnarray}
where $m_{p} = 0.938$ GeV is a proton mass. Taking into account
(\ref{32}) and (\ref{37}) we find
\begin{equation}\label{40}
    \Omega_{B} = \Omega_{qm}\,{\sqrt{\frac{3 \pi}{2 g}}}
    \ \frac{m_{p}}{m_{Pl}}
    \left(1 - \mbox{e}^{-\Gamma_{\phi}\, \Delta t} \right),
\end{equation}
where $g = G\,m^{2}$ is the gravitational coupling constant for
the quantum $\phi$ with mass $m$. Since we suppose that
$\Gamma_{\phi} \sim g$, the density $\Omega_{B}$ is the function
on the coupling constant $g$. This function vanishes at $g = 0$
and tends to zero as $g^{-1/2}$ at $g \rightarrow \infty$. It has
one maximum. Let us fix the coupling constant $g$ by maximum value
of $\Omega_{B}(g)$. Then we obtain
\begin{equation}\label{41}
    \Gamma_{\phi}\, \Delta t = 1.256\,.
\end{equation}
For $\Delta t = 14$ Gyr it gives
\begin{equation}\label{42}
    \Gamma_{\phi} = 2.840 \times 10^{-18}\,\mbox{s}^{-1}.
\end{equation}
This rate satisfies inequality (\ref{36}). In addition
\begin{equation}\label{43}
    \Gamma_{\phi} > H_{0},
\end{equation}
where $H_{0} = 71\,\mbox{km}\,\mbox{s}^{-1}\,\mbox{Mpc}^{-1}$ is
the present-day value of the Hubble expansion rate \cite{3}. This
condition means that on average at least one interaction has
occurred over the lifetime of our universe.

\subsection*{\normalsize 4.3 The model of $\Gamma_{\phi}$}
Since the final product of decay of the quantum $\phi$ through the
intermediate creation of neutron is proton, in a first
approximation one can write a simple expression for the mean
probability of the decay $\Gamma_{\phi}$
\begin{equation}\label{44}
    \Gamma_{\phi} = \alpha\,g\,\Delta m\,,
\end{equation}
where $\alpha$ is a fine-structure constant, $\Delta m = 1.293$
MeV is a difference in masses of neutron and proton. Using the
numerical values of the parameters from (\ref{42}) and (\ref{44})
we find
\begin{equation}\label{45}
    g = 19.8 \times 10^{-38}.
\end{equation}
Then the density of baryons is
\begin{equation}\label{46}
    \Omega_{B} = 0.131,
\end{equation}
and the mass of matter quantum $\phi$ corresponding to such
coupling constant is equal to $m = 5.433$ GeV. The density of the
non-baryonic dark matter is
\begin{equation}\label{47}
    \Omega_{CDM} = \Omega_{B} \, \frac{m_{\phi'}}{m_{p}}\,,
\end{equation}
where, according to (\ref{34}), $m_{\phi'} = m - m_{n} - m_{\nu}'
- Q$ is a mass of the quantum $\phi'$, $m_{\nu}' \equiv
m_{\nu}^{2}/(2\,p_{\nu})$, $m_{\nu}$ and $p_{\nu}$ are the
neutrino mass and momentum, $Q$ is the energy of the relative
motion of all particles. Since the contribution from
$\Omega_{CDM}$ to the matter density $\Omega_{M}$ of our universe
is not at least smaller than $\Omega_{B}$ \cite{2,3,4}, then the
mass $m_{\phi'}$ can possess the values within the limits
\begin{equation}\label{48}
    0.938\,\mbox{GeV} \leq m_{\phi'} \leq 4.493\,\mbox{GeV}
\end{equation}
depending on the value of $Q$. Such particles are
non-rela\-ti\-vi\-stic and non-baryonic dark matter is classified
as cold. Hence we find
\begin{equation}\label{49}
    0.131 \leq \Omega_{CDM} \leq 0.627\,.
\end{equation}
The total matter density $\Omega_{M} = \Omega_{B} + \Omega_{CDM}$
can possess any values within the limits
\begin{equation}\label{50}
    0.262 \leq \Omega_{M} \leq 0.758\,.
\end{equation}
We neglect the contribution from neutrino to $\Omega_{M}$ assuming
that $m_{\nu}' \ll m_{n}$. A spread in theoretical values of the
densities $\Omega_{CDM}$ and $\Omega_{M}$ is caused by the fact
that the energy $Q$ is an undefined parameter of the theory. In
principle it can be fixed from the astrophysical data for
$\Omega_{M}$. With the regard for this remark one can compare
(\ref{50}) with the matter density of our universe,
$\Omega_{M}^{obs} = 0.3 \pm 0.1$ \cite{3}, where a spread in
values is related with inaccuracy of measurements.

According to (\ref{33}) and (\ref{50}) the residual density
$\Omega_{X} = \Omega_{qm} - \Omega_{M}$ can possess the values
within the limits
\begin{equation}\label{51}
    0.302 \leq \Omega_{X} \leq 0.798\,.
\end{equation}
The observations in our universe give the restriction
$\Omega_{X}^{obs} = 0.7 \pm 0.1$ \cite{3}. These values agree with
(\ref{51}).

In our approach according to (\ref{31}), (\ref{32}), (\ref{39}),
and (\ref{47}) the residual density $\Omega_{X}$ has only
dynamical nature and it can be attributed to the optically dark
(nonluminous) energy.

The present-day density of optically bright and dark baryons in
our universe consistent with the big-bang nucleosynthesis is
estimated as $\Omega_{B}^{obs} = 0.039 \pm 0.004$ \cite{3}. This
value does not contradict with (\ref{46}), since the latter
determines the maximum possible baryon matter density. The
observed value of the cold dark matter density $\Omega_{CDM}^{obs}
\sim 0.3$ falls within the limits of its theoretical counterpart
(\ref{49}). According to (\ref{39}) the density of the optically
bright baryons in our approach can be estimated as
\begin{equation}\label{52}
    \Omega_{*} = \frac{1}{16}\,\Omega_{B}\,.
\end{equation}
This relation agrees with the data of astronomical observations
which indicate that the baryons in stars account for about 10\% of
all baryons \cite{3,4}.

Then according to (\ref{46}) the contribution from the optically
bright baryons must not exceed
\begin{equation}\label{53}
    \Omega_{*} \sim 0.008\,.
\end{equation}
In this sense this value is in agreement with the observations of
the bright stars, gaseous content of galaxies, groups and
clusters, $\Omega_{*}^{obs} = 3^{+1}_{-2} \times 10^{-3}$
\cite{3,Sal1,Sal2}.

In our approach the densities $\Omega_{M}$ (\ref{50}) and
$\Omega_{X}$ (\ref{51}) are of the same order. Hence the known
problem of present-day coincidence between dark matter and dark
energy components in our universe \cite{24} is solved
automatically.

Thus the model of matter quantum decay rate (\ref{44}) gives the
values of the energy density components in the universe close (on
the order of magnitude) to observed. Let us note that the possible
mass (\ref{48}) of particle of the non-baryonic dark matter lies
in the range of baryon resonances and mesons. Any known particle
here will not do for the role of quantum $\phi'$.

It is interesting that the range of masses (\ref{48}) is close to
one of the possible limits on the mass of the light gluino,
$m_{\widetilde{g}} \lesssim 5$ GeV \cite{3}. However, since the
gluino is the colour octet Majorana fermion partner of the gluon,
it also will not fit for the role of non-baryonic dark matter
particle $\phi'$.

One can solve an inverse problem. Namely, using the observed value
$\Omega_{*}^{obs}$, it is possible to restore the coupling
constant $g$ and then find the masses $m$, $m_{\phi'}$ and
densities $\Omega_{B}$, $\Omega_{*}$, $\Omega_{M}$ and
$\Omega_{X}$.

\subsection*{\normalsize 4.4 The inverse problem}
Let us set $\Omega_{*} = 0.0025$. Fixing $g$ by the maximum of the
function $\Omega_{B}(g)$ as above, we find
\begin{equation}\label{54}
    g = 212 \times 10^{-38}, \quad m = 17.78\,\mbox{GeV}.
\end{equation}
The corresponding values of mass of the quantum $\phi'$ lie in the
range
\begin{equation}\label{55}
    0.938\,\mbox{GeV} \leq m_{\phi'} \leq 16.84\,\mbox{GeV},
\end{equation}
with mean value $\overline{m}_{\phi'} = 8.889$ GeV, for possible
energies of the decay of matter quantum in the domain $0 \leq Q
\leq 15.90$ GeV. According to (\ref{52}) the baryon density is
equal to $\Omega_{B} = 0.04$. This value practically coincides
with $\Omega_{B}^{obs}$. Repeating the calculations such as above,
we obtain the following limits on the components of energy density
for the parameters (\ref{54})
\begin{equation}\label{56}
  0.04 \leq \Omega_{CDM} \leq 0.718,
\end{equation}
with mean value $\overline{\Omega}_{CDM} = 0.379$,
\begin{equation}\label{57}
  0.08 \leq \Omega_{M} \leq 0.758,
\end{equation}
with mean value $\overline{\Omega}_{M} = 0.419$,
\begin{equation}\label{58}
  0.302 \leq \Omega_{X} \leq 0.98,
\end{equation}
with mean value $\overline{\Omega}_{X} = 0.641$, and $\Omega_{DB}
= 0.0375$ for dark baryons, $\Omega_{DB} = \Omega_{B} -
\Omega_{*}$. All these arithmetic mean values are in good
agreement with corresponding experimental data given above.

The lightest supersymmetric particles can be considered as the
candidates for the quantum $\phi'$ of non-baryonic cold dark
matter. Some accelerator experiments give the bound
$m_{\widetilde{\chi}_{1}^{0}} > 10.9$ GeV for the mass of the
lightest stable neutralino $\widetilde{\chi}_{1}^{0}$ \cite{25}
which falls into the range of $m_{\phi'}$ (\ref{55}).

\section*{\normalsize 5. Concluding remarks}
The mechanism of origin of mass-energy constituents of the
universe by means of the collective excitations of the primordial
scalar field, which we propose, can explain the fact that our
universe appears to be populated exclusively with matter than
antimatter.

For the mass $m_{\phi'} \sim 9$ GeV, the energy $Q \sim 8$ GeV is
released in the decay (\ref{34}). Such kinetic energy corresponds
to the temperature $\sim 10^{14}$ K. This temperature can be in
the primary plasma under thermal equilibrium in the early universe
at $\Delta t \sim 10^{-8}$ s. In early, very dense universe among
the particles of primary hot plasma in addition to primordial
protons, electrons and $\nu\,\bar{\nu}$-pairs there must be other
particle-antiparticle pairs with masses $< m_{p}$, secondary
neutrons and photons. In this case the number of light particles
will exceed the number of baryons (cp. \cite{22}). Such very hot
and dense universe will expand and cool down in accordance with
standard big-bang model (see e.g. \cite{3,22}). The questions
concerning initial heating of the early universe and transition to
the radiation dominated phase in our approach exceed the aims of
this paper and require a separate investigation.

Another interesting question which we shall mention here is
connected with the cosmological constant in our universe. Using
the obtained value of the energy density $\Omega_{v}$ (\ref{33})
for the critical density of the universe $\rho_{c} = 1.879 \times
10^{-29}\,h^{2}\,\mbox{g}\,\mbox{cm}^{-3}$ \cite{3} we find the
energy density of small quantum fluctuations of the scalar field
about equilibrium vacuum state $\phi_{vac}$
\begin{equation}\label{59}
    \rho_{v} = 1.125 \times
    10^{-31}\,h^{2}\,\mbox{g}\,\mbox{cm}^{-3}.
\end{equation}
Such density forms the cosmological constant in the universe
\begin{equation}\label{60}
    \Lambda = 2.097 \times
    10^{-58}\,h^{2}\,\mbox{cm}^{-2}.
\end{equation}
If one suppose that the cosmological constant in our universe
arises due to the residual (dark) energy $\Omega_{X}$ \cite{4}
then for $\Omega_{X} = 0.7$ its value will be
\begin{equation}\label{61}
    \Lambda_{X} = 2.454 \times
    10^{-56}\,h^{2}\,\mbox{cm}^{-2}.
\end{equation}
In this case one can neglect the contribution from $\rho_{v}$ to
the total vacuum energy density.

\end{document}